\documentclass[conference]{IEEEtran}
\usepackage{amsmath,amsfonts}
\usepackage{algorithmic}
\usepackage{algorithm}
\usepackage{array}
\usepackage[caption=false,font=footnotesize,labelfont=rm,textfont=rm]{subfig}
\usepackage{textcomp}
\usepackage{stfloats}
\usepackage{url}
\usepackage{verbatim}
\usepackage{graphicx}
\usepackage{cite}
\usepackage{amsmath}
\usepackage{amssymb}
\usepackage{bm}
\usepackage{color}
\allowdisplaybreaks[3]
\begin{document}

\title{Modeling and Optimization for Rotatable Antenna
Enabled Wireless Communication\vspace{-0.3cm}}

\author{\IEEEauthorblockN{
Qingjie Wu\IEEEauthorrefmark{1},
Beixiong Zheng\IEEEauthorrefmark{1}, Tiantian Ma\IEEEauthorrefmark{1}, and Rui Zhang\IEEEauthorrefmark{2}\IEEEauthorrefmark{3}}
\IEEEauthorblockA{\IEEEauthorrefmark{1}School of Microelectronics, South China University of Technology, Guangzhou 511442, China\\
\IEEEauthorrefmark{2}School of Science and Engineering, The Chinese University of Hong Kong, Shenzhen 518172, China\\
\IEEEauthorrefmark{3}Department of Electrical and Computer Engineering, National University of Singapore, Singapore 117583}
\IEEEauthorblockA{Email: miqjwu@mail.scut.edu.cn; bxzheng@scut.edu.cn; mitiantianma@mail.scut.edu.cn; rzhang@cuhk.edu.cn}\vspace{-0.5cm}}

\maketitle

\begin{abstract}
Fluid antenna system (FAS)/movable antenna (MA) has emerged as a promising technology to fully exploit the spatial degrees of freedom (DoFs). In this paper, we propose a new rotatable antenna (RA) model, as a simplified implementation of six-dimensional movable antenna (6DMA), to improve the performance of wireless communication systems. Different from conventional fixed antenna, the proposed RA system can independently and flexibly change the three-dimensional (3D) orientation/boresight of each antenna by adjusting its deflection angles to achieve desired channel realizations. Specifically, we study an RA-enabled uplink communication system, where the receive beamforming and the deflection angles of all RAs are jointly optimized to maximize the minimum signal-to-interference-plus-noise ratio (SINR) among all the users. In the special single-user and free-space propagation setup, the optimal deflection angles are derived in closed form with the maximum-ratio combining (MRC) beamformer applied at the base station (BS). In the general multi-user and multi-path setup, we propose an alternating optimization (AO) algorithm to alternately optimize the receive beamforming and the deflection angles in an iterative manner. Simulation results are provided to demonstrate that the proposed RA-enabled system can significantly outperform other benchmark schemes.
\end{abstract}

\begin{IEEEkeywords}
Rotatable antenna (RA), fluid antenna system (FAS), movable antenna (MA), 6DMA, near-field modeling, pointing vector optimization.
\end{IEEEkeywords}

\section{Introduction}
In the rapidly evolving landscape of global information and communications technology (ICT), the forthcoming sixth-generation (6G) wireless network is envisioned to support even more densely-connected users and devices across more diverse applications and services, thus demanding superior communication and sensing performance beyond the current fifth-generation (5G) wireless network~\cite{ref_6G}. To further improve transmission rate and spectrum efficiency, wireless network tends to adopt drastically more antennas and more advanced multiple-input-multiple-output (MIMO) technologies at both the base stations (BSs) and user/device terminals~\cite{ref_XLMIMO}.

Although larger-scale MIMO can offer more substantial array and spatial multiplexing gains, it comes at the expense of higher hardware cost and power consumption. Furthermore, increasing the number of antennas cannot fully exploit the spatial degrees of freedom (DoFs) since the traditional fixed antennas cannot change their positions or orientations flexibly once deployed. To overcome this limitation, fluid antenna system (FAS)/movable antenna (MA) has been proposed to enable the local movement of antennas in a specified region through different antenna movement mechanisms~\cite{ref_FAS,ref_FAMA,ref_FASTUT,ref_MA_model,ref_MA_Mag}. As compared to fixed antenna, FAS/MA can proactively transform the wireless channels into a more favorable condition and thus achieve higher capacity without increasing the number of antennas. However, FAS/MA mainly adjusts the positions of antennas with their rotations fixed, and its practical implementation is constrained by the response time or movement speed of the antennas. Recently, six-dimensional movable antenna (6DMA) has been proposed to flexibly adjust both the three-dimensional (3D) position and 3D rotation of distributed antennas/arrays based on the user spatial distribution~\cite{ref_6DMA_cont,ref_6DMA_Mag}.

Motivated by the above, we propose in this paper a new antenna architecture, namely rotatable antenna (RA), as a simplified implementation of 6DMA to improve the performance of wireless communications. In the RA system, the deflection angles of each antenna/array can be independently adjusted to change its 3D orientation/boresight while its 3D position is kept constant to reduce the hardware cost and time/energy overhead for antenna position changes in 6DMA. Under this new model, we investigate an RA-enabled uplink multi-user communication system. Specifically, we aim to maximize the minimum signal-to-interference-plus-noise ratio (SINR) among all the users by jointly optimizing the receive beamforming and the deflection angles of all RAs. For the special single-user and free-space propagation setup, the maximum-ratio combining (MRC) beamformer is applied and the optimal deflection angles are derived in closed form to maximize the channel power gain. For the general multi-user and multi-path setup, an alternating optimization (AO) algorithm is proposed to alternately optimize the receive beamforming and the deflection angles in an iterative way. In particular, the subproblem that optimizes the deflection angles is transformed into a pointing vector optimization problem, which is then solved by the successive convex optimization (SCA) technique to obtain a high-quality suboptimal solution with low complexity. Simulation results show that our proposed RA-enabled system with optimized deflection angles can achieve a much higher SINR as compared to various benchmark schemes.

\section{System Model and Problem Formulation}
\begin{figure}[!t]  \centering
	\includegraphics[width=3in]{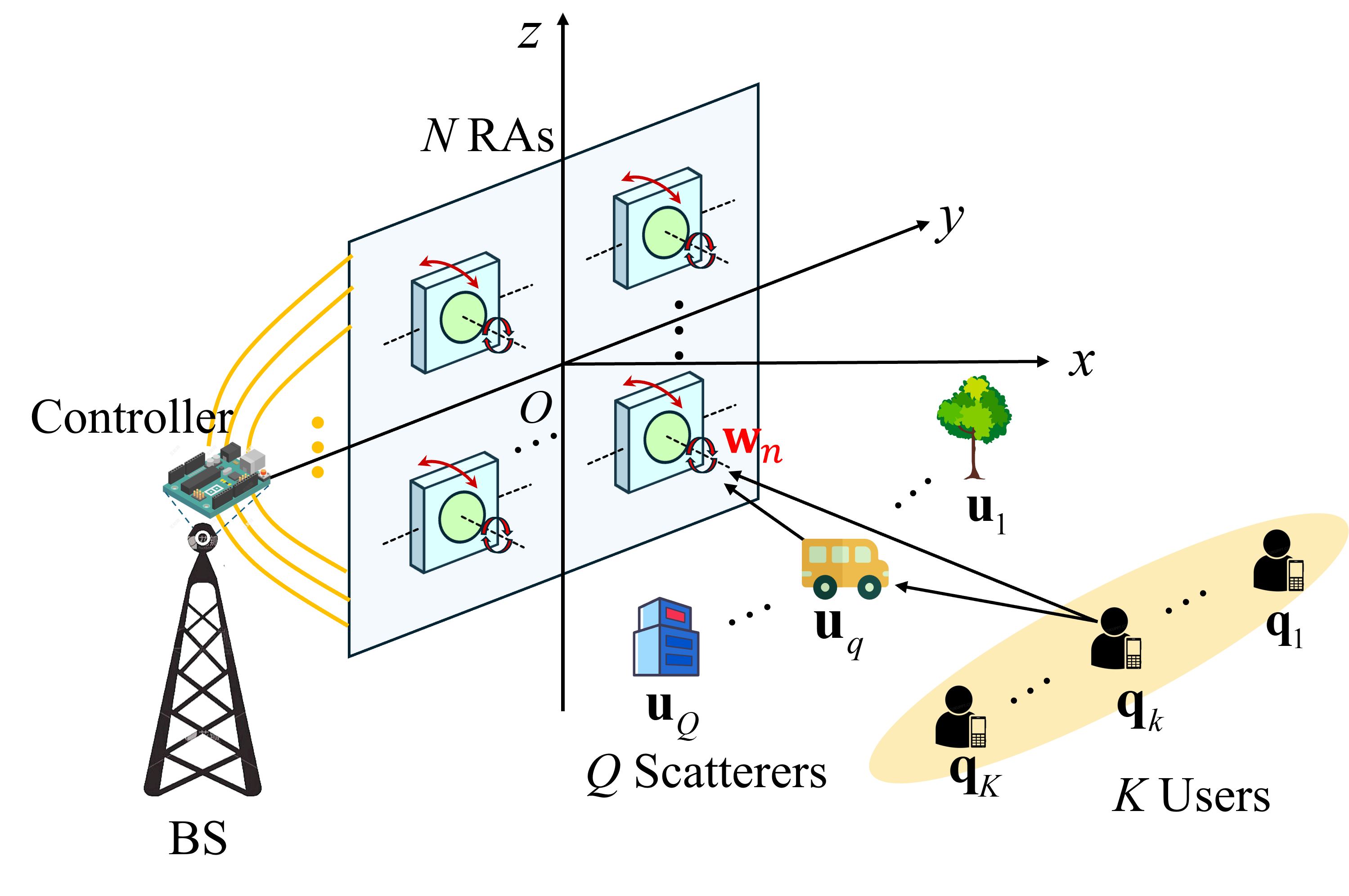}
	\caption{An RA-enabled uplink communication system.}
	\label{fig_system}
\end{figure}

\begin{figure}[!t]
	\centering
	\subfloat[Deflection angles]{
		\includegraphics[width=1.7in]{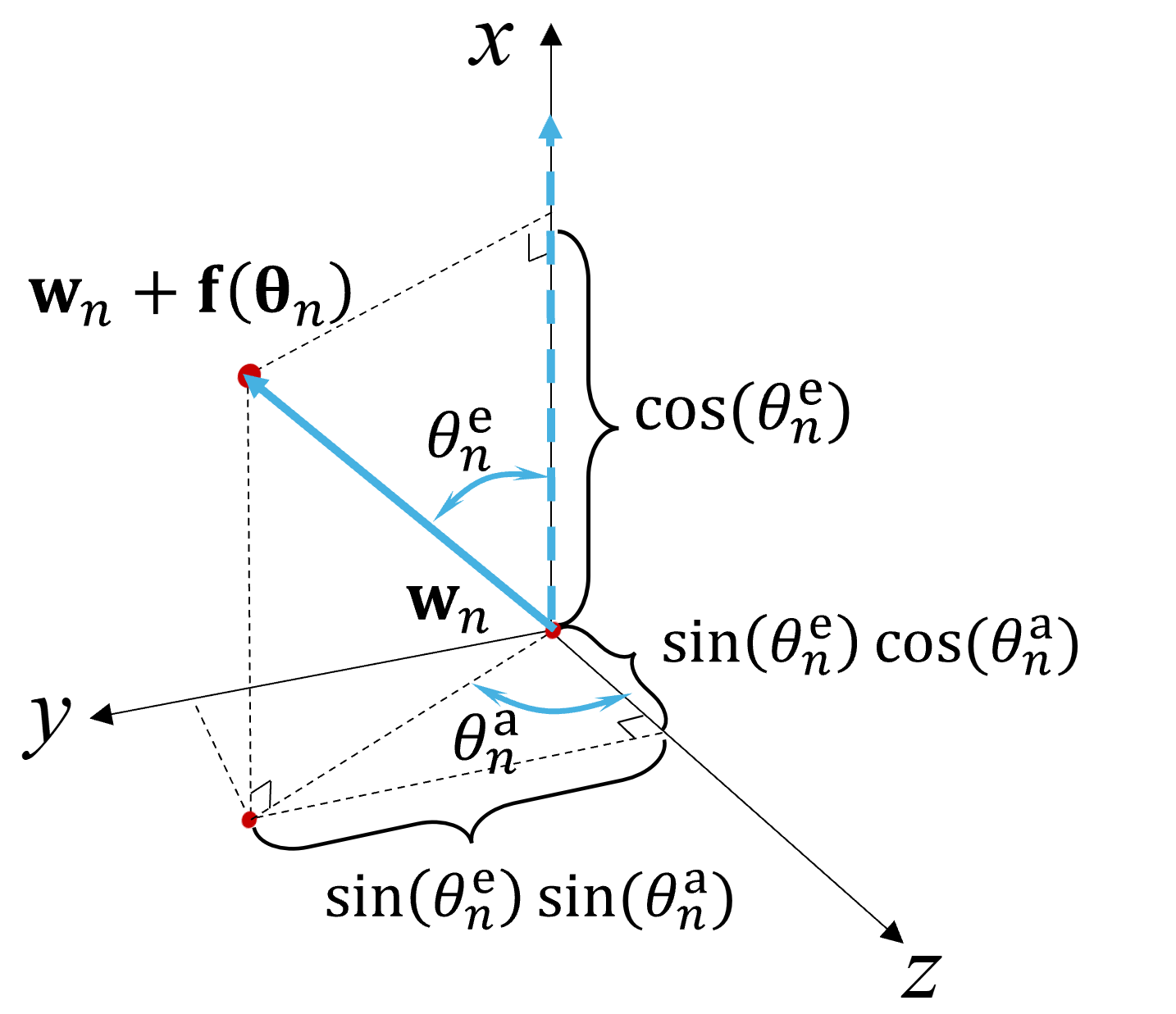}}\hspace{0.8cm}
	\subfloat[Gain pattern]{
		\includegraphics[width=1.2in]{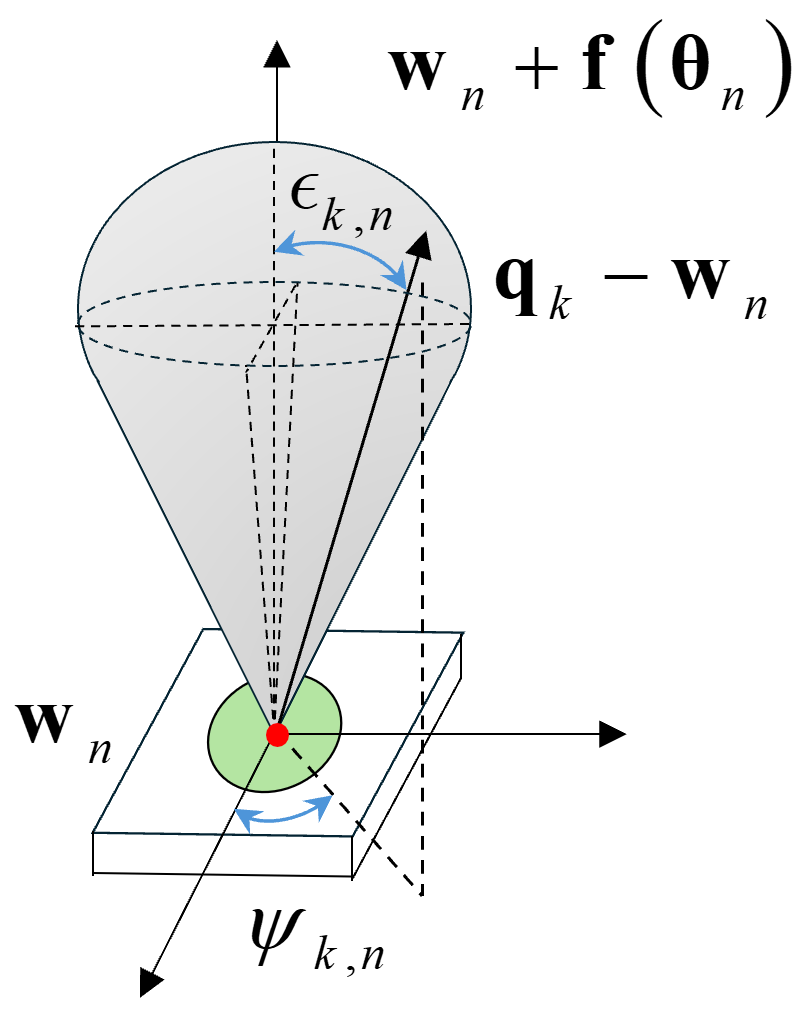}}
	\caption{Illustration of deflection angles and gain pattern of RA~$n$.}
	\label{fig_angle}
\end{figure}

As shown in Fig. \ref{fig_system}, we consider an RA-enabled uplink communication system, where $K$ users (each equipped with a single isotropic fixed antenna) simultaneously transmit their signals in the same time-frequency resource block to a BS equipped with a planar array at fixed position consisting of $N$ directional RAs. Without loss of generality, we assume that the planar array is placed on the $y$-$z$ plane of a 3D Cartesian coordinate system, and the reference positions of RA~$n$ and user~$k$ are represented as $\mathbf{w}_n \in \mathbb{R}^{3\times 1}$ and $\mathbf{q}_k \in \mathbb{R}^{3\times 1}$, respectively. The reference orientations/boresights of all RAs are assumed to be parallel to the $x$-axis, and the orientation/boresight of each RA can be independently adjusted in 3D direction mechanically and/or electrically through a common smart controller. As shown in Fig. \ref{fig_angle}(a), the 3D orientation/boresight adjustment of each RA can be described by two deflection angles, i.e., the eccentric angle and the azimuth angle. Specifically, for RA~$n$, the eccentric angle $\theta_n^{\mathrm{e}}$ represents the angle between the norm vector of RA~$n$ and the $x$-axis (the reference orientation/boresight), while the azimuth angle $\theta_n^{\mathrm{a}}$ is the angle between the projection of the norm vector of RA~$n$ onto the $y$-$z$ plane and the $z$-axis. To characterize the 3D orientation/boresight of each RA, we define a pointing vector corresponding to the eccentric angle $\theta_n^{\mathrm{e}}$ and azimuth angle $\theta_n^{\mathrm{a}}$ of RA~$n$ as
\begin{equation}
	\label{deqn_ex1a}
	\mathbf{f}(\bm{\theta}_n) = [\mathrm{cos}(\theta_n^{\mathrm{e}}),\mathrm{sin}(\theta_n^{\mathrm{e}})\mathrm{sin}(\theta_n^{\mathrm{a}}),\mathrm{sin}(\theta_n^{\mathrm{e}})\mathrm{cos}(\theta_n^{\mathrm{a}})]^T,
\end{equation}
where $\bm{\theta}_n \triangleq [\theta_n^{\mathrm{e}},\theta_n^{\mathrm{a}}]^T$ is defined as the deflection angle vector of RA~$n$, and $\|\mathbf{f}(\bm{\theta}_n)\| = 1$. Due to the maneuverability restriction of the motor and to avoid antenna coupling between any two RAs, the eccentric angle of each RA should be within a specific range, i.e.,
\begin{align}
	\label{deqn_ex2a}
	0 \leq \theta_n^{\mathrm{e}} \leq \theta_{\mathrm{max}}, \forall n,
\end{align}
where $\theta_{\mathrm{max}} \in [0,\frac{\pi}{2}]$ is the maximum eccentric angle that each RA is allowed to adjust.

%\begin{figure}[!t]  \centering
%	\includegraphics[width=2in]{fig_architecture}
%	\caption{RA architecture.}
%	\label{fig_architecture}
%\end{figure}

The effective antenna gain for each RA depends on both signal arrival/departure angle and antenna directional gain pattern, which characterizes the antenna radiation power distribution over different directions. In this paper, we consider a generic directional gain pattern for each RA as follows~\cite{ref_Antenna}.
\begin{equation}
	\label{deqn_ex3a}
	G_e(\epsilon,\psi) = \left\{\begin{matrix}
		G_0 \mathrm{cos}^{2p} (\epsilon), & \epsilon \in [0,\frac{\pi}{2}), \psi \in [0,2\pi) \\
		0,\quad\quad\quad\quad\quad & \mathrm{otherwise},\quad\quad\quad\quad\quad\quad
	\end{matrix}\right.
\end{equation}
where $\epsilon$ and $\psi$ are the elevation and azimuth angles of any spatial point with respect to the RA's current orientation/boresight as shown in Fig. \ref{fig_angle}(b), $p$ determines the directivity of antenna, and $G_0$ is the maximum gain in the boresight direction (i.e., $\epsilon = 0$) with $G_0 = 2(2p + 1)$ to satisfy the law of power conservation. 

We consider a scattering environment with $Q$ distributed scatterers, where the position of scatterer~$q$ is represented by $\mathbf{u}_q \in \mathbb{R}^{3\times 1}$. Based on the \textit{Friis Transmission Equation} and directional gain pattern adopted in \eqref{deqn_ex3a}, and assuming the gain of the transmit antenna at each user is $G_t = 1$, the channel power gain between user~$k$ and RA~$n$ and that between scatterer~$q$ and RA~$n$ can be respectively modeled as~\cite{ref_Friis}
\begin{align}
	g_{k,n}(\bm{\theta}_n) & = \left(\frac{\lambda}{4\pi r_{k,n}}\right)^2 G_0 \mathrm{cos}^{2p} (\epsilon_{k,n}), \label{deqn_ex4a}\\
	m_{q,n}(\bm{\theta}_n) & = \left(\frac{\lambda}{4\pi d_{q,n}}\right)^2 G_0 \mathrm{cos}^{2p} (\tilde{\epsilon}_{q,n}), \label{deqn_ex5a}
\end{align}
where $\lambda$ is the signal wavelength, $r_{k,n} = \|\mathbf{q}_k - \mathbf{w}_n\|$ and $d_{q,n} = \|\mathbf{u}_q - \mathbf{w}_n\|$ are the distances between user~$k$ and RA~$n$ and that between scatterer~$q$ and RA~$n$, respectively, $\mathrm{cos}(\epsilon_{k,n}) \triangleq \mathbf{f}^T(\bm{\theta}_n) \vec{\mathbf{q}}_{k,n}$ and $\mathrm{cos}(\tilde{\epsilon}_{q,n}) \triangleq \mathbf{f}^T(\bm{\theta}_n) \vec{\mathbf{u}}_{q,n}$ are the projection between the user~$k$'s direction vector $\vec{\mathbf{q}}_{k,n} \triangleq \frac{\mathbf{q}_k - \mathbf{w}_n}{\|\mathbf{q}_k - \mathbf{w}_n\|}$ and the pointing vector of RA~$n$ and that between the scatterer~$q$'s direction vector $\vec{\mathbf{u}}_{q,n} \triangleq \frac{\mathbf{u}_q - \mathbf{w}_n}{\|\mathbf{u}_q - \mathbf{w}_n\|}$ and the pointing vector of RA~$n$, respectively.
%\begin{align}
%	\mathrm{cos}(\epsilon_{k,n}) & = \frac{\mathbf{f}^T(\bm{\theta}_n)(\mathbf{q}_k - \mathbf{w}_n)}{\|\mathbf{q}_k - \mathbf{w}_n\|}, \label{deqn_ex6a}\\
%	\mathrm{cos}(\tilde{\epsilon}_{q,n}) & = \frac{\mathbf{f}^T(\bm{\theta}_n)(\mathbf{u}_q - \mathbf{w}_n)}{\|\mathbf{u}_q - \mathbf{w}_n\|}. \label{deqn_ex7a}
%\end{align}

For the multi-path channel between RA~$n$ and user~$k$, by considering the geometric propagation, the line-of-sight (LoS) channel component $h_{k,n}^{\mathrm{LoS}}(\bm{\theta}_n)$ and the non-LoS (NLoS) channel component $h_{k,n}^{\mathrm{NLoS}}(\bm{\theta}_n)$ can be separately modeled by~\cite{ref_multipath_LoS,ref_multipath_NLoS}
\begin{align}
	h_{k,n}^{\mathrm{LoS}}(\bm{\theta}_n) & = \sqrt{g_{k,n}(\bm{\theta}_n)} e^{-j\frac{2\pi}{\lambda}r_{k,n}}, \label{deqn_ex8a}\\
	h_{k,n}^{\mathrm{NLoS}}(\bm{\theta}_n) & = \sum_{q=1}^{Q}{\frac{\sqrt{\sigma_q m_{q,n}(\bm{\theta}_n)}}{t_{k,q}}} e^{-j\frac{2\pi}{\lambda}(d_{q,n}+t_{k,q})+j\phi_q}, \label{deqn_ex9a}
\end{align}
where $\sigma_q$ represents the radar cross section (RCS) of scatterer~$q$, modeled as an independent and identically distributed (i.i.d.) positive random variable, $\phi_q$ represents the phase shift introduced by scatterer~$q$, which is modeled as an i.i.d. random variable with uniform distribution over $[-\pi,\pi)$, and $t_{k,q} = \|\mathbf{q}_k - \mathbf{u}_q\|$ denotes the distance between user~$k$ and scatterer~$q$. Thus, by superimposing the LoS and NLoS channel components, the overall multi-path channel between user~$k$ and the BS is given by
\begin{equation}
	\label{deqn_ex10a}
	\mathbf{h}_k(\bm{\Theta}) = \mathbf{h}_k^{\mathrm{LoS}}(\bm{\Theta}) + \mathbf{h}_k^{\mathrm{NLoS}}(\bm{\Theta}),
\end{equation}
where $\bm{\Theta} \triangleq [\bm{\theta}_1,\bm{\theta}_2,\dots,\bm{\theta}_N]$ is the deflection angle matrix of all RAs, $\mathbf{h}_k^{\mathrm{LoS}}(\bm{\Theta}) \triangleq [h_{k,1}^{\mathrm{LoS}}(\bm{\theta}_1),h_{k,2}^{\mathrm{LoS}}(\bm{\theta}_2),\dots,h_{k,N}^{\mathrm{LoS}}(\bm{\theta}_N)]^T$ and $\mathbf{h}_k^{\mathrm{NLoS}}(\bm{\Theta}) \triangleq [h_{k,1}^{\mathrm{NLoS}}(\bm{\theta}_1),h_{k,2}^{\mathrm{NLoS}}(\bm{\theta}_2),\dots,h_{k,N}^{\mathrm{NLoS}}(\bm{\theta}_N)]^T$ are the LoS and NLoS channel components between user~$k$ and the BS, respectively.

For the uplink communication, the received signal at the BS can be expressed as
\begin{equation}
	\label{deqn_ex11a}
	\mathbf{y} = \sum_{k=1}^{K}{\mathbf{h}_k(\bm{\Theta}) \sqrt{P_k}s_k} + \mathbf{n},
\end{equation}
where $P_k$ and $s_k$ are the transmit power and information-bearing signal of user~$k$, respectively, and $\mathbf{n}$ is the additive white Gaussian noise (AWGN), following the zero-mean circularly symmetric complex Gaussian (CSCG) distribution with power $\sigma^2$, i.e., $\mathbf{n} \sim \mathcal{CN}(0,\sigma^2\mathbf{I}_N)$. Upon receiving $\mathbf{y}$, the BS applies a linear receive beamforming vector $\mathbf{v}_k^H\in \mathbb{C}^{1\times N}$ with $\|\mathbf{v}_k\| = 1$ to extract the signal of user~$k$, i.e.,
\begin{equation}
	\label{deqn_ex12a}
	y_k = \mathbf{v}_k^{H} \mathbf{h}_k(\bm{\Theta}) \sqrt{P_k}s_k + \sum_{j=1,j\ne k}^{K}{\mathbf{v}_k^{H} \mathbf{h}_j(\bm{\Theta}) \sqrt{P_j}s_j}+ \mathbf{v}_k^{H}\mathbf{n}.
\end{equation}
Accordingly, the SINR for decoding the information from user~$k$ is given by
\begin{equation}
	\label{deqn_ex13a}
	\gamma_k = \frac{\bar{P}_k |\mathbf{v}_k^{H} \mathbf{h}_k(\bm{\Theta})|^2}{\sum_{j=1,j\ne k}^{K}{\bar{P}_j|\mathbf{v}_k^{H} \mathbf{h}_j(\bm{\Theta})|^2}+1},
\end{equation}
where $\bar{P}_k = \frac{P_k}{\sigma^2}$ denotes the transmit signal-to-noise ratio (SNR) for user~$k$.

In this paper, we aim to maximize the minimum SINR among all the users by jointly optimizing the receive beamforming matrix $\mathbf{V}\triangleq [\mathbf{v}_1,\mathbf{v}_2,\dots,\mathbf{v}_K]$ and deflection angle matrix $\bm{\Theta}$ of all RAs, subject to their eccentric angle constraints in \eqref{deqn_ex2a}. Thus, the optimization problem is formulated as\footnote{To obtain the theoretical performance gain brought by the RA, we assume that the channel state information (CSI) of all channels involved is perfectly known at the BS.}
\begin{subequations}\label{eq:1}
  \begin{alignat}{2}
    \mathrm{(P1)} \quad \max_{\mathbf{V},\bm{\Theta}} \quad & \min_{k}\gamma_k & \label{eq:1A}\\
    \mbox{s.t.} \quad 
    & 0\leq \theta_n^{\mathrm{e}} \leq \theta_{\mathrm{max}}, \forall n, \label{eq:1B}\\
    & \|\mathbf{v}_k\| = 1, \forall k. \label{eq:1C}
  \end{alignat}
\end{subequations}

\section{Single-User System with Free-Space Propagation}
In this section, we consider the single-user and free-space propagation setup, i.e., $K = 1$ and $Q = 0$, to draw essential insight into (P1). Thus, the channel modeled in \eqref{deqn_ex10a} reduces to $\mathbf{h}_1(\bm{\Theta}) = \mathbf{h}_1^{\mathrm{LoS}}(\bm{\Theta})$. In this case, no inter-user interference (IUI) is present, and thus (P1) is simplified to (by dropping the user index)
\begin{subequations}\label{eq:2}
	\begin{alignat}{2}
		\mathrm{(P2)} \quad \max_{\mathbf{v},\bm{\Theta}} \quad & \bar{P} |\mathbf{v}^{H} \mathbf{h}^{\mathrm{LoS}}(\bm{\Theta})|^2 & \label{eq:2A}\\
		\mbox{s.t.} \quad 
		& \eqref{eq:1B}, \eqref{eq:1C}. \label{eq:2B}
	\end{alignat}
\end{subequations}
For any given deflection angle matrix $\bm{\Theta}$ in the single-user case, it is known that the MRC beamformer is the optimal receive beamforming solution to problem (P2)~\cite{ref_MRC}, i.e., $\mathbf{v}^{\star} = \frac{\mathbf{h}^{\mathrm{LoS}}(\bm{\Theta})}{\|\mathbf{h}^{\mathrm{LoS}}(\bm{\Theta})\|}$. Then, substituting $\mathbf{v}^{\star}$ into \eqref{eq:2A} yields the following SNR expression,
\begin{equation}
	\label{deqn_ex1b}
	\gamma = \bar{P} \|\mathbf{h}^{\mathrm{LoS}}(\bm{\Theta})\|^2 = \frac{\bar{P} G_0 \lambda^2}{16\pi^2} \sum_{n=1}^{N}{\frac{\mathrm{cos}^{2p} (\epsilon_n)}{r_n^2}}.
\end{equation}
Exploiting the result in \eqref{deqn_ex1b}, problem (P2) can be decomposed into $N$ subproblems, each of which independently optimizes the deflection angle vector of one RA. For RA~$n$, the corresponding subproblem is given by
\begin{subequations}\label{eq:3}
	\begin{alignat}{2}
		\mathrm{(P3)} \quad \max_{\bm{\theta}_n} \quad & \left(\mathbf{f}^T(\bm{\theta}_n) \vec{\mathbf{q}}_n \right)^{2p} & \label{eq:3A}\\
		\mbox{s.t.} \quad 
		& 0\leq \theta_n^{\mathrm{e}} \leq \theta_{\mathrm{max}}, \label{eq:3B}
	\end{alignat}
\end{subequations}
where the constant term is omitted in \eqref{eq:3A}. Since $p \geq 0$, by maximizing the projection between unit vectors $\mathbf{f}(\bm{\theta}_n)$ and $\vec{\mathbf{q}}_n$, the optimal solution to problem (P3) is obtained as
\begin{subequations}\label{deqn_ex2b}
	\begin{align}
	    \theta_n^{\mathrm{a}\star} & = 	\mathrm{arctan2}\left({\vec{\mathbf{q}}_n^T \mathbf{e}_2},{\vec{\mathbf{q}}_n^T \mathbf{e}_3} \right), \label{deqn_ex2b1}\\
		\theta_n^{\mathrm{e}\star} & = \mathrm{min} 	\left \{\mathrm{arccos}\left(\vec{\mathbf{q}}_n^T \mathbf{e}_1\right),\theta_{\mathrm{max}} \right \}, \label{deqn_ex2b2}
	\end{align}
\end{subequations}
where $\mathbf{e}_1 = [1,0,0]^T$, $\mathbf{e}_2 = [0,1,0]^T$, and $\mathbf{e}_3 = [0,0,1]^T$.

According to the optimal deflection angles in \eqref{deqn_ex2b}, we can find that each RA prefers to adjust its deflection angles to tune its orientation/boresight toward the user. This is expected since the user can achieve the maximum antenna gain $G_0$ when the boresight direction of RA is aligned with user direction, i.e., $\mathbf{f}(\bm{\theta}_n) = \frac{\mathbf{q} - \mathbf{w}_n}{\|\mathbf{q} - \mathbf{w}_n\|}$. Therefore, by relaxing the eccentric angle constraint \eqref{eq:3B}, the upper bound for user's SNR is given by
\begin{equation}
	\label{deqn_ex3b}
	\gamma_{\mathrm{ub}} = \frac{\bar{P} G_0 \lambda^2}{16\pi^2} \sum_{n=1}^{N}{\frac{1}{\|\mathbf{w}_n - \mathbf{q}\|^2}}.
\end{equation}

\section{Multi-User System under Multi-path Channel}
In this section, we consider the general multi-user and multi-path setup, i.e., $K > 1$ and $Q\geq 0$. Specifically, an AO algorithm is proposed to optimize the receive beamforming and deflection angles iteratively.

\subsection{Receive Beamforming Optimization}
For given deflection angle matrix $\mathbf{\Theta}$, the channel from user~$k$ to the BS modeled in \eqref{deqn_ex10a} is fixed, and problem (P1) reduces to (by simplifying $\mathbf{h}_k(\bm{\Theta})$ to $\mathbf{h}_k$)
\begin{subequations}\label{eq:4}
	\begin{alignat}{2}
		\mathrm{(P4)} \quad \max_{\mathbf{V}} \quad & \min_{k} \frac{\bar{P}_k |\mathbf{v}_k^{H} \mathbf{h}_k|^2}{\sum_{j=1,j\ne k}^{K}{\bar{P}_j|\mathbf{v}_k^{H} \mathbf{h}_j|^2}+1} & \label{eq:4A}\\
		\mbox{s.t.} \quad 
		& \eqref{eq:1C}. \label{eq:4B}
	\end{alignat}
\end{subequations}
In the multi-user case, the zero-forcing (ZF) and minimum mean-square error (MMSE) beamforming are two commonly used linear receivers to suppress IUI and enhance the SINR for each user, described as follows.

\textit{1) ZF Beamforming:} The ZF receiver is designed to completely remove the IUI, which requires $N \geq K$. For user~$k$, the ZF receive beamforming, denoted by $\mathbf{v}_k^{\mathrm{ZF}}$, should satisfy $(\mathbf{v}_k^{\mathrm{ZF}})^H \bar{\mathbf{H}}_k = \mathbf{0}_{1\times (K-1)}$ with $\bar{\mathbf{H}}_k \triangleq [\mathbf{h}_1,\dots,\mathbf{h}_{k-1},\mathbf{h}_{k+1},\dots,\mathbf{h}_K]$. Therefore, the ZF receive beamforming for user~$k$ is expressed as
\begin{equation}
	\label{deqn_ex1c}
	\mathbf{v}_k^{\mathrm{ZF}} = \frac{(\mathbf{I}_N - \bar{\mathbf{H}}_k(\bar{\mathbf{H}}_k^H \bar{\mathbf{H}}_k)^{-1}\bar{\mathbf{H}}_k^H) \mathbf{h}_k}{\|(\mathbf{I}_N - \bar{\mathbf{H}}_k(\bar{\mathbf{H}}_k^H \bar{\mathbf{H}}_k)^{-1}\bar{\mathbf{H}}_k^H) \mathbf{h}_k\|}, \forall k,
\end{equation}
where $\mathbf{I}_N - \bar{\mathbf{H}}_k(\bar{\mathbf{H}}_k^H \bar{\mathbf{H}}_k)^{-1}\bar{\mathbf{H}}_k^H$ is the projection matrix into the space orthogonal to the columns of $\bar{\mathbf{H}}_k$.

\textit{2) MMSE Beamforming:} The SINR in \eqref{deqn_ex13a} is a generalized Rayleigh quotient with respect to $\mathbf{v}_k$, and the receive SINR for each user can be maximized by the MMSE beamforming~\cite{ref_MMSE}. Thus, the optimal solution to problem (P4) can be obtained as
\begin{equation}
	\label{deqn_ex2c}
	\mathbf{v}_k^{\mathrm{MMSE}} = \frac{\mathbf{C}_k^{-1} \mathbf{h}_k}{\|\mathbf{C}_k^{-1} \mathbf{h}_k\|}, \forall k,
\end{equation}
where $\mathbf{C}_k \triangleq \sum_{j=1,j\ne k}^{K}{\bar{P}_j \mathbf{h}_j \mathbf{h}_j^H} + \mathbf{I}_N$ is the interference-plus-noise covariance matrix. To reduce the dimension of matrix inversion from $N\times N$ to $(K - 1)\times(K - 1)$, by applying the \textit{Woodbury matrix identity}, $\mathbf{C}_k^{-1}$ is equivalently expressed as
\begin{align}
	\label{deqn_ex3c}
	\mathbf{C}_k^{-1} = & (\mathbf{I}_N + \bar{\mathbf{H}}_k \mathbf{P}_k \bar{\mathbf{H}}_k^{H})^{-1} \nonumber\\
	= & \mathbf{I}_N - \bar{\mathbf{H}}_k(\mathbf{P}_k^{-1} + \bar{\mathbf{H}}_k^H\bar{\mathbf{H}}_k)^{-1} \bar{\mathbf{H}}_k^H,
\end{align}
where $\mathbf{P}_k \triangleq \mathrm{diag}(\bar{P}_1,\dots,\bar{P}_{k-1},\bar{P}_{k+1},\dots,\bar{P}_K)$.

\subsection{deflection Angle Optimization}
For given receive beamforming matrix $\mathbf{V}$, by introducing the slack optimization variable $\eta$ to denote the minimum SINR, problem (P1) can be written as
\begin{subequations}\label{eq:5}
	\begin{alignat}{2}
		\mathrm{(P5)} \quad \max_{\eta,\bm{\Theta}} \quad & \eta & \label{eq:5A}\\
		\mbox{s.t.} \quad 
		& \gamma_k \geq \eta, \forall k, \label{eq:5B}\\
		& \eqref{eq:1B}. \label{eq:5C}
	\end{alignat}
\end{subequations}
The above subproblem is challenging to solve since constraint \eqref{eq:5B} is non-convex and the azimuth and eccentric angles, i.e., $\theta_n^{\mathrm{a}}$ and $\theta_n^{\mathrm{e}}$, are coupled in the pointing vector $\mathbf{f}(\bm{\theta}_n)$.

Note that the pointing vector $\mathbf{f}(\bm{\theta}_n)$ is essentially a unit vector on the unit sphere, and the deflection angles mainly affect the channel power gains through the projections $\mathrm{cos}(\epsilon_{k,n})$ and $\mathrm{cos}(\tilde{\epsilon}_{k,n})$ as shown in \eqref{deqn_ex4a} and \eqref{deqn_ex5a}. For the convenience of subsequent derivation, we introduce an auxiliary variable $\mathbf{f}_n \in \mathbb{R}^{3\times 1}$ with $\|\mathbf{f}_n\| = 1$ to equivalently replace the influence of deflection angle vector $\bm{\theta}_n$ on the pointing vector of RA~$n$, i.e., $\mathbf{f}_n \triangleq \mathbf{f}(\bm{\theta}_n)$. Thus, based on \eqref{deqn_ex8a} and \eqref{deqn_ex9a}, the multi-path channel between user~$k$ and RA~$n$ can be rewritten as
\begin{equation}
	\label{deqn_ex1d}
	\tilde{h}_{k,n}(\mathbf{f}_n) = \alpha_{k,n} \left(\mathbf{f}_n^T\vec{\mathbf{q}}_{k,n}\right)^p + \sum_{q=1}^{Q}{\beta_{k,n,q} \left(\mathbf{f}_n^T\vec{\mathbf{u}}_{q,n}\right)^p},
\end{equation}
where $\alpha_{k,n} \triangleq \frac{\lambda \sqrt{G_0} }{4\pi r_{k,n}} e^{-j\frac{2\pi}{\lambda}r_{k,n}}$ and $\beta_{k,n,q} \triangleq \frac{\lambda \sqrt{G_0 \sigma_q}}{4\pi d_{q,n} t_{k,q}} e^{-j\frac{2\pi}{\lambda}(d_{q,n}+t_{k,q})+j\phi_q}$. Let $\mathbf{F} \triangleq [\mathbf{f}_1,\mathbf{f}_2,\dots,\mathbf{f}_N]$ and $\tilde{\mathbf{h}}_k(\mathbf{F}) \triangleq [\tilde{h}_{k,1}(\mathbf{f}_1),\tilde{h}_{k,2}(\mathbf{f}_2),\dots,\tilde{h}_{k,N}(\mathbf{f}_N)]^T$, and problem (P5) can be transformed into
\begin{subequations}\label{eq:6}
	\begin{alignat}{2}
		\mathrm{(P6)} \quad \max_{\eta,\mathbf{F}} \quad & \eta & \label{eq:6A}\\
		\mbox{s.t.} \quad
		& \frac{\bar{P}_k |\mathbf{v}_k^{H} \tilde{\mathbf{h}}_k(\mathbf{F})|^2}{\sum_{j=1,j\ne k}^{K}{\bar{P}_j|\mathbf{v}_k^H \tilde{\mathbf{h}}_j(\mathbf{F})|^2} + 1} \geq \eta, \forall k, \label{eq:6B}\\
		& \mathbf{f}_n^T \mathbf{e}_1 \geq \mathrm{cos}(\theta_\mathrm{max}), \forall n, \label{eq:6C}\\
		& \|\mathbf{f}_n\| = 1, \forall n, \label{eq:6D}
	\end{alignat}
\end{subequations}
where constraint \eqref{eq:6C} is equivalent to \eqref{eq:1B} and constraint \eqref{eq:6D} ensures that $\mathbf{f}_n$ is a unit vector.

To deal with the fractional structure in the left hand side of constraint \eqref{eq:6B}, we take the logarithmic operation on both sides of constraint \eqref{eq:6B}, resulting in an equivalent form for constraint \eqref{eq:6B}, i.e.,
\begin{align}
	\label{deqn_ex2d}
	& \mathrm{ln}\left(\bar{P}_k |\mathbf{v}_k^{H} \tilde{\mathbf{h}}_k(\mathbf{F})|^2\right) \geq \mathrm{ln}(\eta) \nonumber\\
	& \quad \quad + \mathrm{ln}\left(\sum_{j=1,j\ne k}^{K}{\bar{P}_j|\mathbf{v}_k^H \tilde{\mathbf{h}}_j(\mathbf{F})|^2} + 1\right), \forall k,
\end{align}
which is still difficult to handle since $\tilde{h}_{k,n}(\mathbf{f}_n)$ in \eqref{deqn_ex1d} is neither convex nor concave due to the complex coefficients $\alpha_{k,n}$ and $\{\beta_{k,n,q}\}$. To address this challenge, we adopt the SCA technique to approximate constraint \eqref{deqn_ex2d} as a convex constraint and obtain a local optimal solution to problem (P6) in an iterative manner. Without loss of generality, we present the procedure of the $(i+1)$-th iteration and denote the solutions of $\mathbf{F}$ and $\eta$ obtained in the $i$-th iteration by $\mathbf{F}^{(i)}$ and $\eta^{(i)}$, respectively. By using the first-order Taylor expansions at $\{\mathbf{f}_n^{(i)}\}$, $|\mathbf{v}_k^{H} \tilde{\mathbf{h}}_k(\mathbf{F})|^2$ and $\mathrm{ln}\left(\sum_{j=1,j\ne k}^{K}{\bar{P}_j|\mathbf{v}_k^H \tilde{\mathbf{h}}_j(\mathbf{F})|^2} + 1\right)$ in \eqref{deqn_ex2d} can be respectively linearized as $\Lambda_k^{(i+1)}(\mathbf{F})$ and $\Omega_k^{(i+1)}(\mathbf{F})$, shown at the top of the next page, where $\tilde{\mathbf{h}}_{k,n}^{'} \triangleq \frac{\partial \tilde{h}_{k,n}(\mathbf{f}_n^{(i)})}{\partial \mathbf{f}_n^{(i)}} = \tilde{\alpha}_{k,n} \vec{\mathbf{q}}_{k,n} + \sum_{q=1}^{Q}{\tilde{\beta}_{k,n,q} \vec{\mathbf{u}}_{q,n}}$ with $\tilde{\alpha}_{k,n} \triangleq \alpha_{k,n}p ((\mathbf{f}_n^{(i)})^T\vec{\mathbf{q}}_{k,n})^{p-1}$ and $\tilde{\beta}_{k,n,q} \triangleq \beta_{k,n,q}p ((\mathbf{f}_n^{(i)})^T\vec{\mathbf{u}}_{q,n})^{p-1}$. Similarly, an upper bound for $\mathrm{ln}(\eta)$ is obtained as $\Xi^{(i+1)}(\eta) \triangleq \mathrm{ln}(\eta^{(i)}) + \frac{\eta}{\eta^{(i)}} - 1$ by using its first-order Taylor expansions at $\eta^{(i)}$. In this way, constraint \eqref{deqn_ex2d} can be approximated by
\begin{figure*}[ht]
	\begin{align}
		\label{deqn_ex3d}
		\Lambda_k^{(i+1)}(\mathbf{F}) \triangleq |\mathbf{v}_k^{H} \tilde{\mathbf{h}}_k(\mathbf{F}^{(i)})|^2 + \mathrm{Re}\left\{\left(\mathbf{v}_k^{H} \tilde{\mathbf{h}}_k(\mathbf{F}^{(i)}) \right)^{\ast} \sum_{n=1}^{N}{ v_{k,n}^{\ast}\left(\tilde{\mathbf{h}}_{k,n}^{'}\right)^T\left(\mathbf{f}_n - \mathbf{f}_n^{(i)}\right) }\right\}.	
	\end{align}
\end{figure*}
\begin{figure*}[ht]
	\begin{align}
		\label{deqn_ex4d}
		\hspace{-0.4cm}\Omega_k^{(i+1)}(\mathbf{F}) \triangleq 
		\mathrm{ln}\left(\sum_{j=1,j\ne k}^{K}{\bar{P}_j|\mathbf{v}_k^H \tilde{\mathbf{h}}_j(\mathbf{F}^{(i)})|^2} + 1 \right) + \frac{\sum_{j=1,j\ne k}^{K}{\bar{P}_j \mathrm{Re}\left\{\left(\mathbf{v}_k^{H} \tilde{\mathbf{h}}_j(\mathbf{F}^{(i)}) \right)^{\ast} \sum_{n=1}^{N}{ v_{k,n}^{\ast}\left(\tilde{\mathbf{h}}_{j,n}^{'}\right)^T\left(\mathbf{f}_n - \mathbf{f}_n^{(i)}\right)}\right\}}}{\sum_{j=1,j\ne k}^{K}{\bar{P}_j|\mathbf{v}_k^H \tilde{\mathbf{h}}_j(\mathbf{F}^{(i)})|^2} + 1}.	
	\end{align} \hrulefill  
\end{figure*}
\begin{equation}
	\label{deqn_ex5d} \mathrm{ln}\left(\bar{P}_k\Lambda_k^{(i+1)}(\mathbf{F})\right) \geq \Omega_k^{(i+1)}(\mathbf{F}) + \Xi^{(i+1)}(\eta), \forall k.
\end{equation}
Thus, problem (P6) can be approximated by the following problem in the $(i+1)$-th iteration.
\begin{subequations}\label{eq:7}
	\begin{alignat}{2}
		\mathrm{(P7)} \quad \max_{\eta,\mathbf{F}} \quad & \eta & \label{eq:7A}\\
		\mbox{s.t.} \quad 
		& \eqref{eq:6C}, \eqref{eq:6D}, \eqref{deqn_ex5d}. \label{eq:7B}
	\end{alignat}
\end{subequations}
Problem (P7) is still non-convex due to the unit constraint for $\mathbf{f}_n$. For convenience, we first relax the equality constraint \eqref{eq:6D} as $\|\mathbf{f}_n\| \leq 1$, yielding the following problem.
\begin{subequations}\label{eq:8}
	\begin{alignat}{2}
		\mathrm{(P8)} \quad \max_{\eta,\mathbf{F}} \quad & \eta & \label{eq:8A}\\
		\mbox{s.t.} \quad 
		& \|\mathbf{f}_n\| \leq 1, \forall n, \label{eq:8B}\\
		& \eqref{eq:6C}, \eqref{deqn_ex5d}. \label{eq:8C}
	\end{alignat}
\end{subequations}
We can prove that problem (P8) is a convex optimization problem, which can be solved by the CVX solver. Note that the optimal value obtained by problem (P8) serves as an upper bound for that of problem (P7) due to the relaxation of the equality constraint \eqref{eq:6D}.

\subsection{Overall Algorithm}
We summarize the proposed AO algorithm in Algorithm~\ref{alg1}. Since the optimal objective value $\eta$ is non-decreasing over iterations and must be upper-bounded, Algorithm \ref{alg1} is guaranteed to converge. The complexity order of Algorithm~\ref{alg1} is $\mathcal{O}\left(L(KN^3 + N^{3.5} \mathrm{ln}(1/\varepsilon))\right)$, where $L$ denotes the required iteration number for algorithm convergence.

After obtaining the optimal solutions of $\mathbf{V}$ and $\mathbf{F}$ through Algorithm \ref{alg1}, the pointing vector needs to be recovered as a unit vector, i.e., $\mathbf{f}_n^{\star} = \frac{\mathbf{f}_n}{\|\mathbf{f}_n\|}$. Furthermore, the original problem aims to obtain the deflection angle matrix $\bm{\Theta}$, and thus an additional step is needed to transform the optimized pointing vector into the desired deflection angles. For RA~$n$, based on the obtained pointing vector $\mathbf{f}_n^{\star }$, the corresponding deflection angle vector $\bm{\theta}_n^{\star}$ can be obtained as
\begin{subequations}\label{deqn_ex5e}
	\begin{align}
		\theta_n^{\mathrm{a}\star} & = 	\mathrm{arctan2}\left({(\mathbf{f}_n^{\star})^T \mathbf{e}_2},{(\mathbf{f}_n^{\star})^T \mathbf{e}_3} \right), \label{deqn_ex5e1}\\
		\theta_n^{\mathrm{e}\star} & = \mathrm{arccos}\left((\mathbf{f}_n^{\star})^T \mathbf{e}_1 \right). \label{deqn_ex5e2}
	\end{align}
\end{subequations}
As a result, the optimal minimum SINR $\eta^{\star}$ can be calculated by substituting the obtained $\bm{V}^{\star}$ and $\bm{\Theta}^{\star}$ into \eqref{deqn_ex13a}.

\begin{algorithm}[h]
	\caption{Proposed AO Algorithm for Solving (P1).}
	\begin{algorithmic}[1] \label{alg1}
		\STATE Initialize pointing vector $\mathbf{F}^{(0)}$, minimum receive SINR $\eta^{(0)}$, and threshold $\varepsilon > 0$. Let $i=0$.
		\REPEAT
		\STATE Set $i=i+1$.
		\STATE Given $\mathbf{F}^{(i-1)}$, calculate $\mathbf{V}^{(i)}$ according to \eqref{deqn_ex1c}/\eqref{deqn_ex2c}.
		\STATE Given $\mathbf{V}^{(i)}$, $\mathbf{F}^{(i-1)}$ and $\eta^{(i-1)}$, obtain $\mathbf{F}^{(i)}$ and $\eta^{(i)}$ by solving problem (P8).
		\UNTIL $|\frac{\eta^{(i)} - \eta^{(i-1)}}{\eta^{(i-1)}}| \leq \varepsilon$.
		\STATE Output: $\mathbf{V} = \mathbf{V}^{(i)}$ and $\mathbf{F} = \mathbf{F}^{(i)}$.
	\end{algorithmic} 
\end{algorithm} 

\section{Simulation Results}
In this section, we provide numerical results to evaluate the performance of our proposed RA-enabled communication system and optimization algorithm. We assume that the array at the BS is a uniform planar array (UPA) with $N = N_y N_z$ RAs, where $N_y$ and $N_z$ denote the numbers of RAs along $y$- and $z$-axes, respectively, and centered at the origin. For ease of exposition, assuming that $N_y$ and $N_z$ are odd numbers, the position of RA~$n$ is $\mathbf{w}_n = \mathbf{w}_{(n_z-1)N_y+n_y} = [0,n_yd,n_zd]^T$, where $n_y = 0,\pm 1,\dots,\pm \frac{N_y-1}{2}$, $n_z = 0,\pm 1,\dots,\pm \frac{N_z-1}{2}$, and $d = \frac{\lambda}{2}$ is the antenna spacing. Unless otherwise stated, we set $\lambda =$ 0.125 meter (m), $p =$ 4, $\theta_{\mathrm{max}} = \frac{\pi}{6}$, and $N_y = N_z = \bar{N}$.

\subsection{Single-User System}
First, we consider a single-user system under free-space propagation with the transmit SNR $\bar{P} =$ 30 dB, and the user position given by $\mathbf{q} = [r\mathrm{cos}(\varphi),r\mathrm{sin}(\varphi),0]^T$, where $r =$ 50 m is the distance between the user and the origin and $\varphi \in [-\frac{\pi}{2},\frac{\pi}{2}]$ is the azimuth angle of the user. In the following, based on the MRC beamforming, we compare the maximum receive SNRs obtained with the optimal deflection angles (i.e., $\mathbf{\Theta} = \mathbf{\Theta}^{\star}$ given in \eqref{deqn_ex2b}) and reference deflection angles (i.e., $\mathbf{\Theta} = \mathbf{0}$), denoted by $\gamma(\mathbf{\Theta}^{\star})$ and $\gamma(\mathbf{0})$, respectively.

\begin{figure}[!t] \centering
	\includegraphics[width=3in]{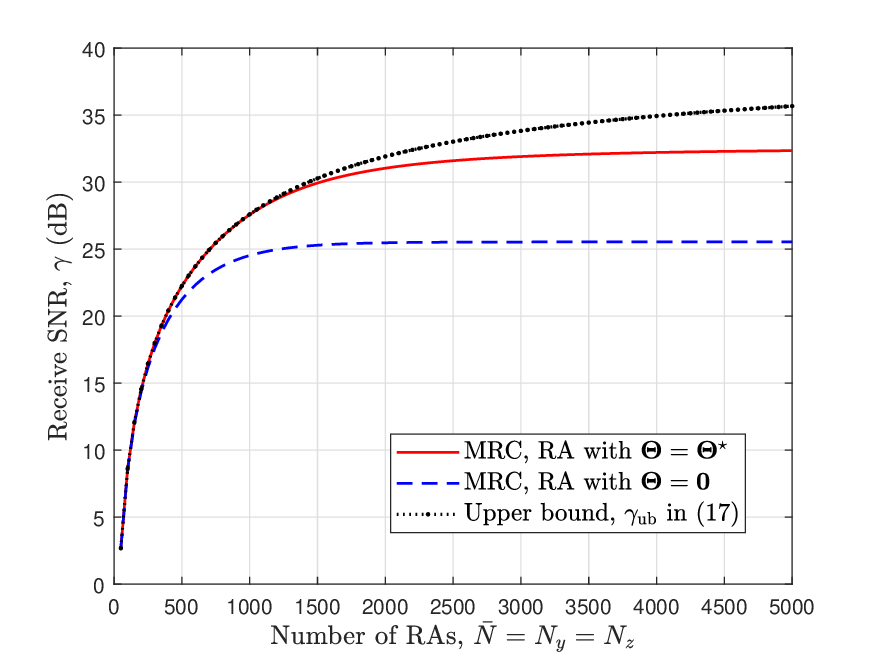}
	\caption{Receive SNRs obtained by different schemes versus the number of RAs $\bar{N} = N_y = N_z$ for the single-user system with $\varphi =$ 0.}
	\label{fig_single_numRA}
\end{figure}

Fig. \ref{fig_single_numRA} plots the receive SNR versus the number of RAs. It is observed that for a moderate $\bar{N}$, the SNRs obtained by all the schemes are close since $\mathbf{\Theta}^{\star} \approx \mathbf{0}$ in the considered case where the user is located directly in front of the UPA when the distance $r$ is much larger than the physical dimension of the UPA. With a further increase in the number of RAs, $\gamma(\mathbf{0})$ and $\gamma(\mathbf{\Theta}^{\star})$ continue to increase and successively reach saturation, and $\gamma(\mathbf{\Theta}^{\star})$ obtains a significantly higher asymptotic value since the RAs located at the edge of the UPA can effectively adjust their 3D orientations/boresights to serve the user with a larger directional gain. This result suggests that large-scale antenna array systems have a greater need for RAs to further improve performance gains.

\begin{figure}[!t] \centering
	\includegraphics[width=3in]{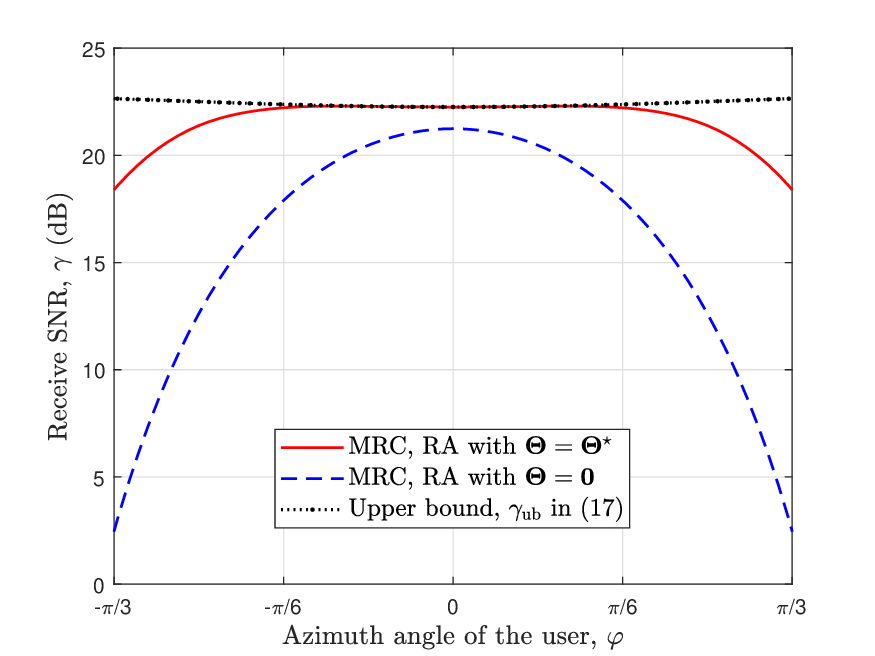}
	\caption{Receive SNRs obtained by different schemes versus the azimuth angle of the user $\psi$ for the single-user system with $N_y$ = $N_z$ = 501.}
	\label{fig_single_angle}
\end{figure}

Fig. \ref{fig_single_angle} shows the receive SNR versus the azimuth angle of the user. As $\varphi$ increases from 0 to $\frac{\pi}{3}$ or decreases from 0 to $-\frac{\pi}{3}$, $\gamma(\mathbf{0})$ drastically decreases since the effective projection between the user direction vector and the pointing vector of each RA decreases, thereby reducing the antennas' directional gains. On the contrary, benefiting from the ability to reorient the boresight direction of RA toward the user to obtain the directional gain, the proposed RA-enabled system can achieve a consistently high SNR even when the user directly faces one of the two ends of the UPA. The above results demonstrate that the deployment of RAs can effectively improve the coverage performance of the BS.

\subsection{Multi-User 
	System}
Next, we consider a multi-user system under the multi-path channel model with $K \geq$ 1 users and $Q =$ 3 scatterers, which respectively  lie evenly on a half circle and randomly in a half circle both centered at the origin with radius $r =$ 50 m. The transmit SNRs of all users are set as the same, i.e., $\bar{P}_k = \bar{P}, \forall k$, and the number of RAs is $N_y = N_z =$ 11. Considering both the ZF and MMSE beamforming, we compare the minimum SINRs (or SNRs in the ZF case) obtained with the optimal deflection angles (i.e., $\mathbf{\Theta} = \mathbf{\Theta}^{\star}$ obtained by Algorithm \ref{alg1}), reference deflection angles (i.e., fixed antenna with $\mathbf{\Theta} = \mathbf{0}$), and random deflection angles (i.e., $\mathbf{\Theta} = \tilde{\mathbf{\Theta}}$, where $\{\tilde{\theta}_{n}^{\mathrm{a}}\}$ and $\{\tilde{\theta}_{n}^{\mathrm{e}}\}$ are randomly generated following a uniform distribution within $[0,2\pi)$ and $[0,\theta_{\mathrm{max}}]$, respectively), denoted by $\eta(\mathbf{\Theta}^{\star})$, $\eta(\mathbf{0})$, and $\eta(\tilde{\mathbf{\Theta}})$, respectively. Furthermore, the isotropic antenna system (let $G_0 =$ 1 and $q =$ 0 in \eqref{deqn_ex4a} and \eqref{deqn_ex5a}) is also considered as a benchmark scheme, and its achieved minimum SINR is denoted by $\eta_{\mathrm{iso}}$.

\begin{figure}[!t]  \centering
	\includegraphics[width=3in]{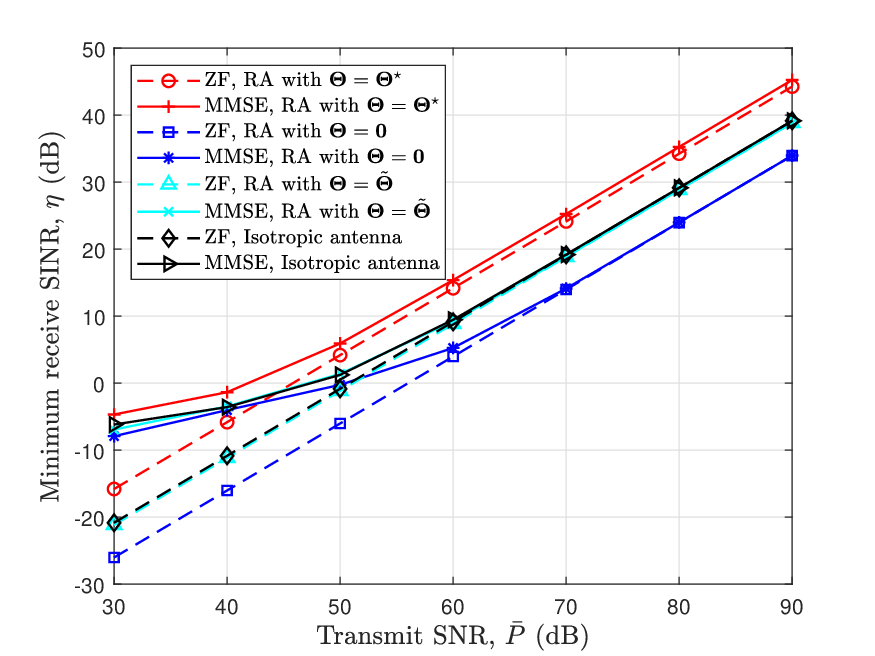} 
	\caption{Minimum receive SINRs obtained by different schemes versus the transmit SNR $\bar{P}$ for the multi-user system with $K =$ 4.}
	\label{fig_multi_power}
\end{figure}

Fig. \ref{fig_multi_power} compares the minimum SINRs obtained by different schemes versus the transmit SNR $\bar{P}$. We observe that with the same linear receiver and transmit power, our proposed RA-enabled system with optimal deflection angles always outperforms other schemes. The minimum SINRs $\eta_{\mathrm{iso}}$ and $\eta(\tilde{\mathbf{\Theta}})$ are similar and always higher than $\eta(\mathbf{0})$. The reason behind this is that the isotropic antennas and the RAs with random deflection angles are statistically equivalent and can both radiate signals in any directions of the BS to serve spatially-distributed users. Furthermore, the MMSE receiver always outperforms the ZF receiver, and the performance gap between them increases as the transmit SNR decreases. This is expected since ZF receiver enhances the noise, and this noise enhancement becomes more pronounced at lower transmit SNR.

\begin{figure}[!t]  \centering
	\includegraphics[width=3in]{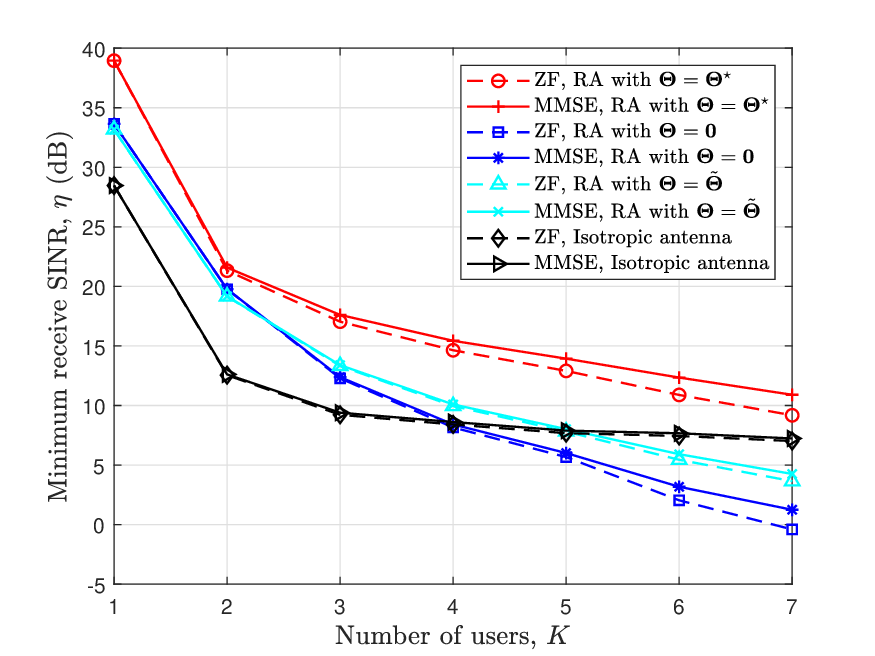} 
	\caption{Minimum receive SINRs obtained by different schemes versus the number of users $K$ for the multi-user system with $\bar{P} = 60$ dB.}
	\label{fig_multi_user}
\end{figure}

Fig. \ref{fig_multi_user} shows the minimum SINRs obtained by different systems versus the number of users $K$. It is observed that the minimum SINRs of all considered schemes decrease as $K$ increases. Particularly, $\eta(\mathbf{\Theta}^{\star})$ is substantially higher compared with the SINRs of other schemes, and their performance gaps increase as the number of users increases. Such performance gains are attributed to our proposed RA-enabled system, which can flexibly adjust the 3D orientations/boresights of all RAs based on the locations of users and scatterers, thus improving the spatial multiplexing gain in multi-user communications.

\section{Conclusions}
In this paper, we proposed a new RA model and aimed to maximize the minimum SINR among all users in an RA-enabled uplink communication system by jointly optimizing the receive beamforming and the deflection angles of all RAs. The optimal closed-form deflection angles were first derived with MRC beamforming applied in the single-user and free-space propagation setup. An AO algorithm was then proposed for the multi-user and multi-path setup, where the ZF/MMSE beamforming was adopted and a pointing vector optimization problem was solved via the SCA technique to obtain a high-quality suboptimal solution to the deflection angles. Simulation results showed that our proposed RA-enabled system with optimized deflection angles can achieve a higher SINR as compared to various benchmark schemes.

\section*{Acknowledgment}
The work of B. Zheng was supported in part by the National Natural Science Foundation of China under Grant 62201214, the Natural Science Foundation of Guangdong Province under Grant 2023A1515011753, the Fundamental Research Funds for the Central Universities under Grant 2024ZYGXZR087, and the GJYC program of Guangzhou under Grant 2024D03J0006. The work of R. Zhang was supported in part by the Guangdong Provincial Key Laboratory of Big Data Computing, the National Natural Science Foundation of China under Grant 62331022, and the Guangdong Major Project of Basic and Applied Basic Research under Grant 2023B0303000001.

\newpage

\vfill


\begin{thebibliography}{1}
\bibliographystyle{IEEEtran}

\bibitem{ref_6G}
C.-X. Wang \textit{et al.}, ``On the road to 6G: Visions, requirements, key technologies, and testbeds,'' \textit{IEEE Commun. Surv. Tutorials}, vol. 25, no. 2, pp. 905–974, 2nd Quart., 2023.

\bibitem{ref_XLMIMO}
Y. Liu \textit{et al.}, ``Near-field communications: A tutorial review,'' \textit{IEEE Open J. Commun. Soc.}, vol. 4, pp. 1999–2049, Aug. 2023.

\bibitem{ref_FAS}
K.-K. Wong, A. Shojaeifard, K.-F. Tong, and Y. Zhang, ``Fluid antenna systems,'' \textit{IEEE Trans. Wireless Commun.}, vol. 20, no. 3, pp. 1950–1962, Mar. 2021.

\bibitem{ref_FAMA}
K.-K. Wong and K.-F. Tong, ``Fluid antenna multiple access,'' \textit{IEEE Trans. Wireless Commun.}, vol. 21, no. 7, pp. 4801–4815, Jul. 2022.

\bibitem{ref_FASTUT}
W. K. New \textit{et al.}, ``A tutorial on fluid antenna system for 6G networks: Encompassing communication theory, optimization methods and hardware designs,'' \textit{arXiv preprint arXiv:2407.03449}, Nov. 2024.

\bibitem{ref_MA_model}
L. Zhu, W. Ma, and R. Zhang, ``Modeling and performance analysis for movable antenna enabled wireless communications,'' \textit{IEEE Trans. Wireless Commun.}, vol. 23, no. 6, pp. 6234–6250, Jun. 2024.

\bibitem{ref_MA_Mag}
L. Zhu, W. Ma, and R. Zhang, ``Movable antennas for wireless communication: Opportunities and challenges,'' \textit{IEEE Commun. Mag.}, vol. 62, no. 6, pp. 114–120, Jun. 2024.

\bibitem{ref_6DMA_cont}
X. Shao, Q. Jiang, and R. Zhang, ``6D movable antenna based on user distribution: Modeling and optimization,'' \textit{IEEE Trans. Wireless Commun.}, vol. 24, no. 1, pp. 355–370, Jan. 2025.

\bibitem{ref_6DMA_Mag}
X. Shao and R. Zhang, ``6DMA enhanced wireless network with flexible
antenna position and rotation: Opportunities and challenges,'' \textit{arXiv preprint arXiv:2406.06064}, Jun. 2024.

\bibitem{ref_Antenna}
C. A. Balanis, \textit{Antenna theory: Analysis and Design.} John wiley \& sons, 2015.

\bibitem{ref_Friis}
H. T. Friis, ``A note on a simple transmission formula,'' \textit{Proceedings of the IRE}, vol. 34, no. 5, pp. 254–256, May. 1946.

\bibitem{ref_multipath_LoS}
Y. Lu and L. Dai, ``Near-field channel estimation in mixed LoS/NLoS environments for extremely large-scale MIMO systems,'' \textit{IEEE Trans. Commun.}, vol. 71, no. 6, pp. 3694–3707, Jun. 2023.

\bibitem{ref_multipath_NLoS}
Z. Dong and Y. Zeng, ``Near-field spatial correlation for extremely large-scale array communications,'' \textit{IEEE Commun. Lett.}, vol. 26, no. 7, pp. 1534–1538, Jul. 2022.

\bibitem{ref_MRC}
D. Tse and P. Viswanath, \textit{Fundamentals of Wireless Communication}. Cambridge, U.K.: Cambridge Univ. Press, 2005.

\bibitem{ref_MMSE}
H. Lu and Y. Zeng, ``Near-field modeling and performance analysis for multi-user extremely large-scale MIMO communication,'' \textit{IEEE Commun. Lett.}, vol. 26, no. 2, pp. 277–281, Feb. 2022.

\end{thebibliography}
\end{document}